\documentclass[aps,prb,twocolumn,showpacs,groupedaddress,longbibliography]{revtex4-1}
\usepackage{amsbsy,amssymb,amsmath,bm}
\usepackage{graphicx}
\usepackage{braket}
\usepackage{float}
\usepackage{textcomp}
\usepackage{color}
\usepackage[colorlinks=true,linkcolor=red]{hyperref}
\usepackage[sort&compress]{natbib}

\newcommand*\diff{\mathop{}\!\mathrm{d}}

\input{epsf}

\begin{document}

\title {Anderson localization in generalized discrete time quantum walks}
\author{I.  Vakulchyk$^{1,2}$, M. V. Fistul$^{1,3}$, P. Qin$^1$ and S. Flach$^1$}
\affiliation {$^{1}$ Center for Theoretical Physics of Complex Systems, Institute for Basic Science (IBS), Daejeon 34051, Republic of Korea \\
$^{2}$Basic Science Program, Korea University of Science and Technology (UST), Daejeon 34113, Republic of Korea\\
$^{3}$ Russian Quantum Center, National University of Science and Technology "MISIS", 119049 Moscow, Russia
}

\date{\today}

\begin{abstract}
We study Anderson localization in a generalized discrete time quantum walk - a unitary map related to a Floquet driven quantum lattice. It is controlled by a quantum coin matrix
which depends on four angles with the meaning of potential and kinetic energy, and external and internal synthetic flux. Such quantum coins can be engineered with microwave pulses in qubit chains.
The ordered case yields a two-band eigenvalue structure on the unit circle which becomes completely flat in the limit of vanishing kinetic energy.
Disorder in the external magnetic field does not impact localization. Disorder in all the remaining angles yields Anderson localization. In particular,
kinetic energy disorder leads to logarithmic divergence of the localization length at spectral symmetry points. Strong disorder in potential and internal magnetic field energies allows
to obtain analytical expressions for spectrally independent localization length which is highly useful for various applications.
\end{abstract}

\maketitle

\section{Introduction}

Quantum random walks were introduced by Aharonov et al \cite{PhysRevA.48.1687} as a counterpart to classical random walks, with the quantum version having much larger path length
due to quantum interference. The quantum computing community developed these concepts substantially e.g. in order to implement a variety of quantum logical elements and protocols
\cite{doi:10.1080/00107151031000110776,Venegas-Andraca2012}.
One should differentiate between so-called continuous time quantum walks, and
discrete time quantum walks (DTQW).
Continuous time quantum walks
are essentially time-dependent Schr\"odinger equations on tight binding networks generated by corresponding
Hamiltonians. Discrete time quantum walks (DTQW) which are the subject of the present work. DTQW are unitary maps on certain networks (graphs). The generating Hamiltonian is not known and needed, although DTQW can be loosely related to the procedure of 
integrating a certain Hamiltonian over a finite time. DTQW realizations are closely related to a number
of concepts in condensed matter physics, including chirality and bulk-boundary correspondence \cite{PhysRevB.88.121406}, novel topological phases \cite{PhysRevB.92.045424},
and the impact of disorder on the DTQW dynamics \cite{PhysRevE.82.031122,PhysRevB.84.195139,PhysRevA.92.052311,PhysRevA.94.023601}. DTQW were implemented in numerous experimental setups, including quantum optical systems \cite{peruzzo2010quantum}, ion traps \cite{schmitz2009quantum} and nuclear magnetic resonance systems \cite{du2003experimental}.

The DTQW has two ingredients - a quantum coin, and a shift (register) operation. So far, quantum coins were chosen
mainly from single parameter (angle) operator distributions, including the well known case of the Hadamard coin \cite{aharonov2001quantum, tregenna2003controlling}. However, the most general quantum coin belongs to a four parameter (angle) family of operators, as shown below. It has been shown \cite{chandrashekar2008optimizing} that generalized coins allow for improved control and optimization. Such a general coin can be implemented in an optical setup utilizing beam splitters \cite{cerf1998optical}, and some parameters may be controlled in other setups \cite{di2004cavity}. We will study the impact of disorder in any of the angles
on the DTQW dynamics. We find novel Anderson localized phases, and rigorously derive scaling relations for weak and strong disorder, and close to symmetry related values in the spectrum.
DTQW turn out to be versatile machines not only for quantum computing, but also as model systems for condensed matter problems, which can be efficiently addressed by avoiding a number of
computational troubles known from Hamiltonian dynamics.

The paper is organized as follows: in section \ref{II} we present the model, elaborate the quantum-mechanical dynamic equations, calculate the dispersion relationship $\omega(k)$ for a generic type of discrete time quantum walk, provide a symmetry analysis of dynamic equations
and introduce a transfer matrix approach for a discrete time quantum walk.
In section \ref{III} we present numerical results on the localization length dependence on the model parameters. In section \ref{IV} we perform analytical derivations of the localization length
 in the limit of weak and strong disorder, and at symmetry points in the spectrum. We discuss and conclude in section \ref{V}.

\begin{figure}[tbp]
\includegraphics[width=0.95 \columnwidth]{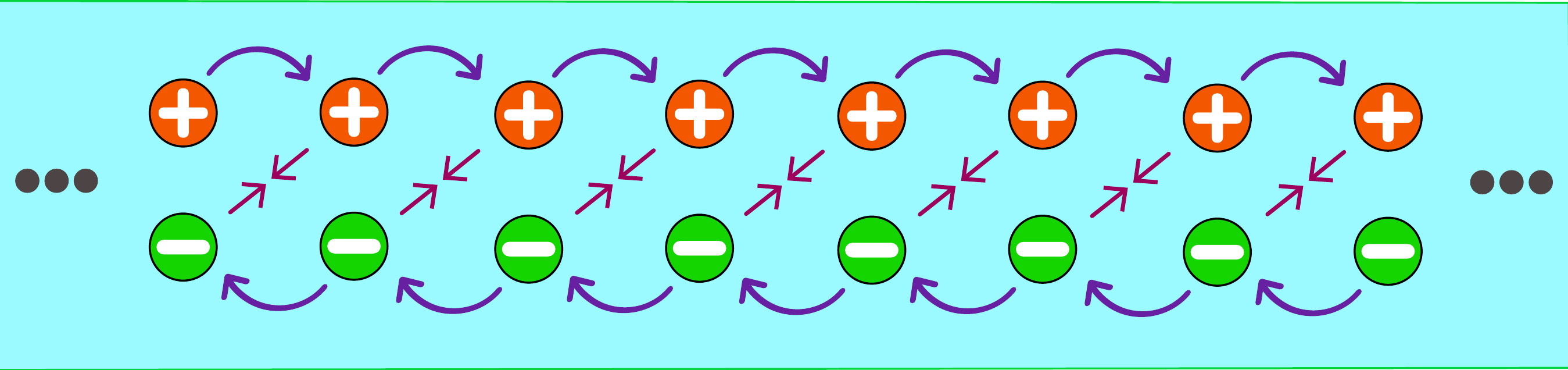}
\caption{ The schematics of the discrete-time quantum walk. The arrows indicate the directions of a single step transfer.
}
\label{fig1}
\end{figure}

\section{Model}
\label{II}
Let us consider a single quantum particle with an internal spin degree of freedom, moving on a one-dimensional lattice. The dynamics of the quantum particle is characterized by a time- and lattice site-dependent \textit{two-component wave function}.
Assume that it evolves under the influence of some periodic Floquet drive. Then, its evolution can be mapped onto a sequence of unitary maps.
As a result, the components of the quantum particle wave function transfer to the right or to the left, and \textit{the quantum-mechanical amplitudes} of such hopping are determined by quantum
coin operators acting independently on each site (Fig.\ref{fig1}).

The dynamics of a quantum particle is characterized by a two-component wave function, $\hat \psi_n (t)=\{\psi_{+,n},\psi_{-,n} \}$, which is defined at discrete times ($t$) and on lattice sites ($n$).
A single site coin operator $\hat U$ is a general unitary matrix of rank 2:
\begin{equation}\label{coin_operator}
  \begin{split}
    \hat U &
=
\begin{pmatrix}
       a& b  \\
       c &  d
       \end{pmatrix}
     =e^{i \varphi}
        \begin{pmatrix}
        e^{i \varphi_1} \cos{\theta} & e^{i \varphi_2} \sin{\theta} \\
        -e^{-i \varphi_2} \sin{\theta} & e^{-i \varphi_1} \cos{\theta}
        \end{pmatrix}. \\
  \end{split}
\end{equation}
A generic coin operator is completely determined by four angles $\varphi, \varphi_{1}, \varphi_2$ and $\theta$.
As it will become evident below, they can be also related to a potential energy, external and internal synthetic flux, and a kinetic energy respectively.
The coin operator can be implemented as an arbitrary two-level system subject to time-dependent perturbations of different durations \cite{CoinOperImpl,Impl1,Impl2,Impl3}.
The coupling between the coin operators $\hat U_n$ and the quantum particle has a form, $ \hat S = \sum_n \ket{n}\bra{n} \otimes \hat U_n$, where
the angles $\varphi, \varphi_{1}, \varphi_2$ and $\theta$ can vary from site to site.

The transfer operator is defined as
\begin{equation}\label{shift_operator}
  \hat T_\pm = \sum_n \ket{n}\bra{n+1} \otimes \ket{\mp}\bra{\mp} +
  \ket{n}\bra{n-1} \otimes \ket{\pm}\bra{\pm},
\end{equation}
and we will use $T_+$ across the paper. Thus, the discrete-time quantum walk is described as the sequence of successive $\hat S$ and $\hat T_+$ operators. The schematic of such dynamics is shown in Fig.\ref{fig1}, and the equations read
\begin{equation} \label{HoppingEquation}
\hat \psi_n(t+1)=\hat M_+ \hat \psi_{n-1}(t)+\hat M_- \hat \psi_{n+1}(t),
\end{equation}
where the matrices $\hat M_{\pm}$ for the translationally invariant case of identical quantum coins are written explicitly as
 \begin{equation} \label{Mmatrix+}
\hat M_+= \left ( \begin{array}{cc}
e^{i(\varphi_1+\varphi) }\cos \theta&   e^{i(\varphi_2+\varphi) }\sin \theta \\
0& 0\\
\end{array} \right )
\end{equation}
and
 \begin{equation} \label{Mmatrix-}
\hat M_-= \left ( \begin{array}{cc}
0&  0\\
-e^{i(\varphi-\varphi_2) }\sin \theta& e^{i(-\varphi_1+\varphi) }\cos \theta \\
\end{array} \right ) .
\end{equation}
The resulting unitary eigenvalue problem is solved by finding the eigenvectors
$\{ \hat \psi_n \}$ with
$\hat \psi_n(t+1)=e^{-i\omega  }\hat \psi_n(t)$ and the eigenvalues $e^{-i\omega}$, where $\omega$ is the eigenfrequency of the discrete-time quantum walker.

\subsection{Ordered case}

In the absence of spatial disorder 
all coin operators are identical, and the unitary map equations (\ref{HoppingEquation}) are invariant under discrete translations.
The eigenvectors are then given by plane waves $\hat \psi_n=e^{i k n} \hat \psi(k)$, where $k$ is the wave vector, and $ \hat \psi(k)$ is the two-component eigenvector in the Bloch basis (also called
polarization vector). The quantum particle dynamics is then fully determined by the
dispersion relation
\begin{equation} \label{Dispersion}
 \cos{(\omega-\varphi)} = \cos{\theta} \cos{(k-\varphi_1)} \;.
\end{equation}
The spectrum consists of two bands. The polarization vectors are obtained as
\begin{equation}
\label{EVs}
\frac{\psi_{+,k}}{\psi_{-,k}}  = e^{i(\varphi_2 - \varphi_1)} \frac{\cos \theta - e^{i( [\omega(k) - \varphi] -[k -\varphi_1 ] ) }}{ \sin \theta  } \;.
\end{equation}
It follows that $\theta$ is a kinetic energy parameter which controls the width of each band from its maximal value $\pi$ for $\theta=0$ to a dispersionless 
(flat) band with width zero for $\theta=\pi/2$.
The angle $\varphi$ corresponds to a potential energy term which renormalizes the frequency $\omega$. The angle $\varphi_1$ renormalizes the wave number $k$ similar to a flux threading
a large one-dimensional chain with periodic boundary conditions. The angle $\varphi_2$ instead relates to an internal synthetic flux which impacts the phase shift between the two components of the
polarization vector only.

\begin{figure}[tbp]
\includegraphics[width=0.95 \columnwidth]{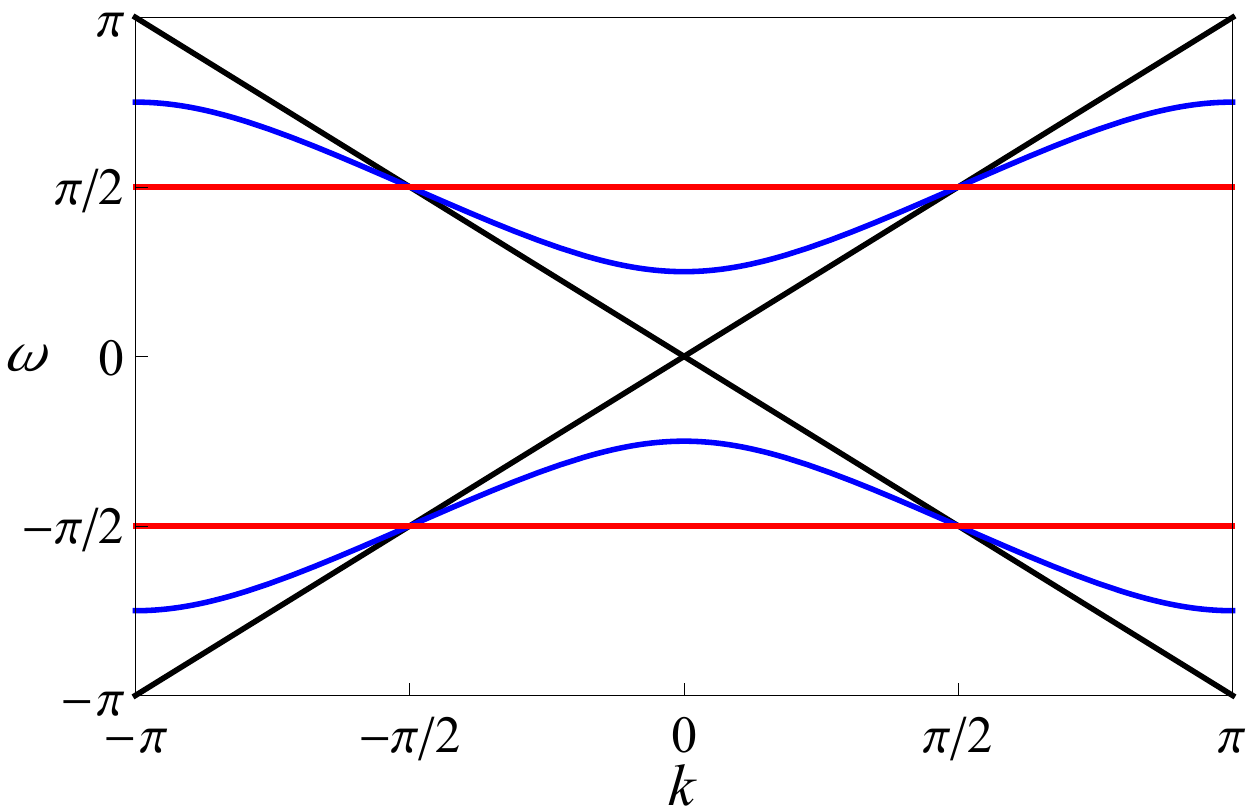}
\caption{ The dispersion relation $\omega(k)$ for different values of $\theta$: $\theta=0 $ (no gap, black solid line), $\pi/4$ (finite gap, blue line), $\pi/2$ (flat 
bands, red lines). Here, $\varphi=\varphi_1=0$. }
\label{fig2}
\end{figure}

For a generic value of $\theta$ the two bands have finite width and are gapped away from each other (e.g. blue lines, $\theta=\pi/4$ in Fig.\ref{fig2}).
For $\theta=0$ the two bands turn into straight lines which cross, leading to a vanishing gap and a one-dimensional Dirac-like cone (black lines in Fig.\ref{fig2}).
Finally for
$\theta = \pi / 2$ the spectrum $\omega(k) = \varphi \pm \pi/2$ consists of \textit{two flat bands} (Fig.\ref{fig2}).
This corresponds to macroscopic degeneracy. Linear combinations of Bloch eigenstates are easily shown to allow for
\textit{compact ($2$-sites) localized states} residing on a pair of neighbouring sites $m$ and $m+1$:
\begin{equation}
\label{CLS}
        \hat \psi_n = \frac{1}{\sqrt{2}} \begin{pmatrix}
            1\\ 0
        \end{pmatrix}
        \delta_{n,m}
        +
 \frac{1}{\sqrt{2}}
        \begin{pmatrix}
            0 \\ ie^{-i\varphi_2}
        \end{pmatrix}
        \delta_{n,m+1}\;.
\end{equation}

\subsection{Symmetries}

For arbitrary unitary coin operators the quantum walk (\ref{HoppingEquation}) possesses \textit{bipartite} or \textit{sublattice symmetry}, since even/odd sites are connected to odd/even sites only.
The bipartite lattice symmetry implies that the spectrum is invariant under frequency shifts $\omega \rightarrow \omega+\pi$ with the following transformation rules for eigenvectors:
\begin{equation}
\{ \omega\; , \; \hat \psi_n \} \rightarrow \{ \omega+\pi \; , \; (-1)^n \hat \psi_n \}
\end{equation}
Note that any arbitrary spatial disorder in the coin operators is preserving the sublattice symmetry.

For site-independent angles $\varphi_n\equiv \varphi$ and $\varphi_{1,n} \equiv \varphi_1$
the quantum walk  (\ref{HoppingEquation}) possesses an additional
 \textit{particle-hole symmetry}, which implies that the spectrum is invariant under frequency shifts $\omega \rightarrow -\omega+2\varphi$ with the following transformation rules for
the eigenvectors:
\begin{widetext}
\begin{equation}
\label{TRS}
\{ \omega\; , \;  \psi^+_n \;,\; \psi^-_n \} \rightarrow
\{ -\omega+2\varphi \; , \; \psi^{+*}_n \exp\left(2i(\varphi_2-\sum_{m=-\infty}^{n}\varphi_{1,m})\right)
 \;,\;
 \psi^{-*}_n \exp\left(-2i\sum_{m=-\infty}^{n-1}\varphi_{1,m})\right) \} \;.
\end{equation}
\end{widetext}

\subsection{Disorder and transfer matrix approach}
We turn to the disordered case where any of the quantum coin angles $(\varphi, \varphi_1, \varphi_2)$ or $\theta$ are assumed to be uncorrelated random functions of the quantum particle position $n$.
In this case the \textit{transfer matrix approach } is useful for both numerical and analytical approaches of computing the localization length. With Eq.(\ref{HoppingEquation}) it follows
\begin{widetext}
 \begin{eqnarray} \label{HoppingeqEX}
e^{-i\omega}\psi_{+,n} = e^{i[\varphi_{1,(n-1)}+\varphi_{n-1}]} \cos \theta_{n-1} \psi_{+,(n-1)} \; - \;
e^{i[\varphi_{2,(n-1)}+\varphi_{n-1}]} \sin \theta_{n-1} \psi_{-,(n-1)} \;, \\
e^{-i\omega}\psi_{-,n} = e^{-i[\varphi_{2,(n+1)}-\varphi_{n+1}]} \sin \theta_{n+1} \psi_{+,(n+1)}
+e^{i[-\varphi_{1,(n+1)}+\varphi_{n+1}]} \cos \theta_{n+1} \psi_{-,(n+1)}\;. \label{HoppingeqEX-2}
\end{eqnarray}
\end{widetext}
The usual transfer matrix for a one-dimensional lattice with two components per lattice site and nearest neighbour coupling is expected to have rank 4. However,
the special structure of the shift operator (\ref{shift_operator}) allows to reduce the transfer matrix rank to 2.
This can be observed with a redefinition of the two component field
$\hat{\Psi}_n=(\psi_{+,(n)};\psi_{-,(n-1)})$, which then leads to the transfer matrix equation
\begin{equation} \label{TransfermatrixEq}
\hat \Psi_n=\hat T_{n-1} \hat \Psi_{n-1},
\end{equation}
where the transfer matrix $T$ has a following form:
 \begin{equation} \label{Trmatrix}
\hat T_n= e^{i\varphi_{1,n}}\left ( \begin{array}{cc}
e^{i\omega +i\varphi_n}\sec \theta_n &  e^{i\varphi_{2,n}}\tan \theta_n \\
e^{-i\varphi_{2,n}}\tan \theta_n & e^{-i\omega -i\varphi_n}\sec \theta_n
\end{array} \right ) .
\end{equation}
It follows that disorder in the external synthetic flux $\varphi_1$ does not lead to the localization of the quantum particle wave function, since such a disorder is only modifying the phase difference on neighbouring sites, while keeping the amplitude ratio unchanged. However, uncorrelated disorder in all other quantum coin angles $\theta_n$ (kinetic energy) $\varphi_n$ (potential energy) and $\varphi_{2,n}$ (internal synthetic flux)
will lead to Anderson localization, as discussed in what follows. We will use probability distribution functions
\begin{equation}
\mathcal{P}(x)=1/(2W) \;,\; x_0 - W \leq x \leq x_0+W
\label{PDF}
\end{equation}
and $\mathcal{P}(x)=0$ elsewhere, where
$x$ stands for any relevant angle, and $x_0$ is the corresponding first moment (average).
The disorder strength $0 \leq W \leq \pi$.

\section{Numerical computation of the localization length}
\label{III}
In this section, we numerically compute the localization length $L_{loc}(\omega)$ using the canonical approach elaborated e.g. in Ref.~\onlinecite{mackinnon1983scaling}. We start with a nonzero  $\hat \Psi_0$ and repeatedly apply randomly generated matrices (\ref{Trmatrix}) to this wave function according to (\ref{TransfermatrixEq}).
We use  $|\hat \Psi_n|=\sqrt{|\Psi_{+,n}|^2+|\Psi_{-,n}|^2}$ and compute the Lyapunov exponent at the $N$th iteration as
\begin{equation} \label{LyapExp}
\lambda_N = \frac{1}{N}\sum_{n=1}^{N} \ln\left(\left|\hat{\Psi}_n\right|\right) \;.
\end{equation}
The localization length is then obtained as\cite{mackinnon1983scaling}
\begin{equation} \label{LocNum}
L_{loc}=1/\lambda_N.
\end{equation}
In order to ensure convergence, we used $N=10^6-10^9$ matrices.
The validity of the approach was cross checked by direct diagonalization for large finite systems.
We further note that for a disorder which is weak as compared to the gap of the band structure of the ordered case, the density of states deep in the gap will be strictly zero.
Nevertheless the transfer matrix approach will generate a certain (finite) localization length, which will correspond to some additional fictious defect state with a corresponding frequency.

\subsection{Disorder in $\varphi$}

\begin{figure}[h]
\includegraphics[width=0.95 \columnwidth]{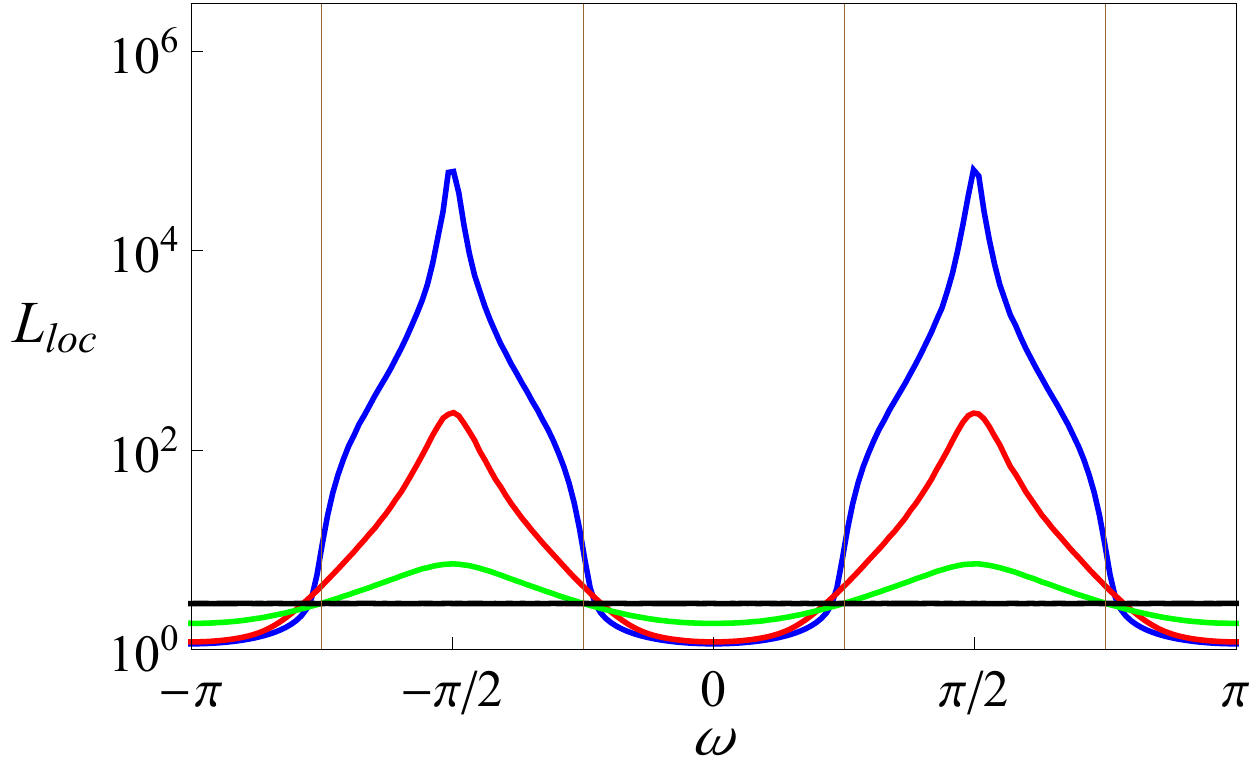}
\caption{The dependence of the localization length on the frequency $\omega$ for disorder in $\varphi$.
From top to bottom at $\omega=\pi/2$: $W=\pi/20,\pi/5,\pi/2,\pi$. Here $\theta=\pi/4$.
The red vertical lines indicate the boundaries of the allowed bands (\ref{Dispersion}).}
\label{fig3}
\end{figure}

We remind that the angle $\varphi$ can be interpreted as a potential energy. The corresponding disorder is similar to {\it diagonal disorder} for tight binding Hamiltonians\cite{mackinnon1983scaling}.
Without loss of generality we can take $x_0=0$ in (\ref{PDF}).
We observe that the localization length $L_{loc}$ is always finite for any finite but nonzero strength of the disorder $W$, see Fig.\ref{fig3}.

For weak disorder $W \ll \pi$ we find that
$L_{loc}$ is large as the frequency $\omega$ is inside the allowed bands of the ordered case (\ref{Dispersion})
and decreases rapidly as the frequency $\omega$ moves inside the gaps, with an anomalous enhancement of $L_{loc}$
at the band centers $\omega = \pm \pi/2$. As the strength of disorder increases the localization length variations diminish, and remarkably $L_{loc}$ becomes
independent of $\omega$ for $W=\pi$.
Variation of $\theta$ does not qualitatively changes the outcome. However, in the special case $\theta = \pm \pi/2$ the localization length vanishes $L_{loc}=0$.
Indeed, the eigenstates are then still compactly localized in full accord with (\ref{CLS}), while the eigenfrequencies simply become $\omega_n=\pm \pi/2 + \varphi_n$.

\subsection{Disorder in $\varphi_2$}

\begin{figure}[h]
\includegraphics[width=0.95 \columnwidth]{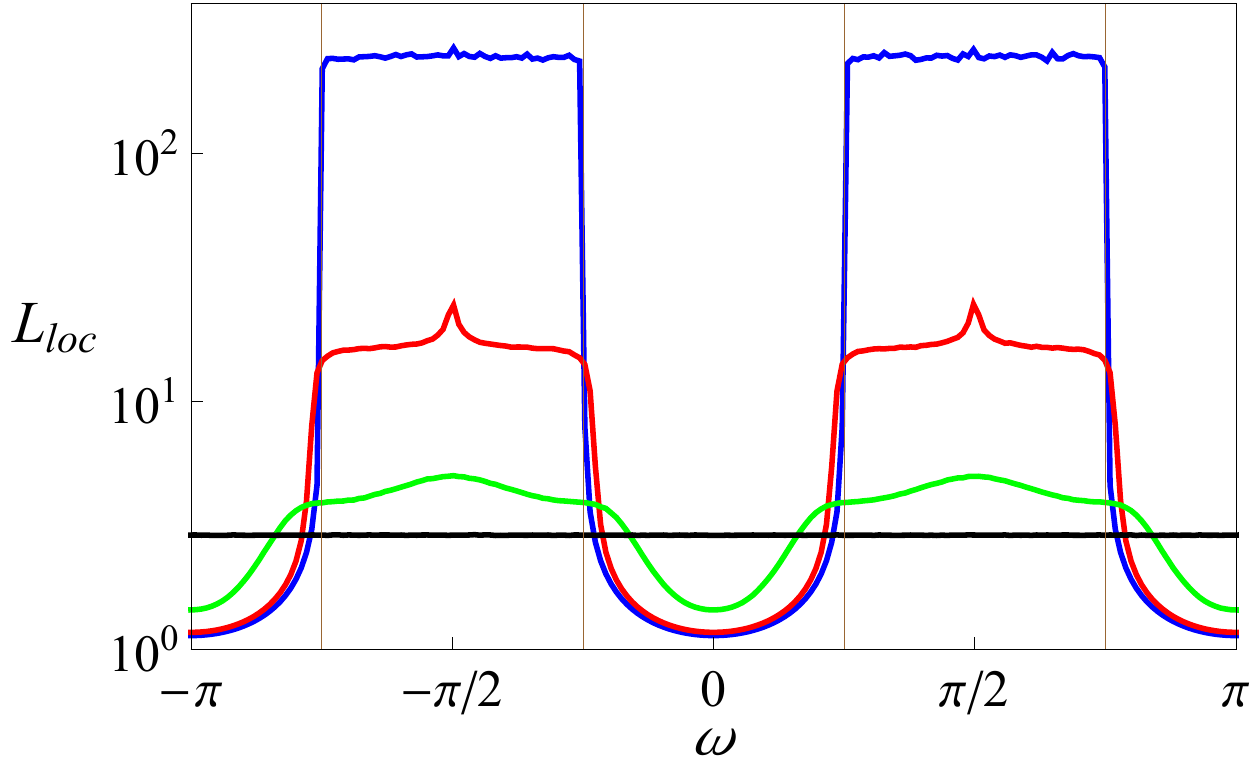}
\caption{
The dependence of the localization length on the frequency $\omega$ for disorder in $\varphi_2$.
From top to bottom at $\omega=\pi/2$: $W=\pi/20,\pi/5,\pi/2,\pi$. Here $\theta=\pi/4$.
The red vertical lines indicate the boundaries of the allowed bands (\ref{Dispersion}).}
\label{fig4}
\end{figure}

\begin{figure}[tbp]
\includegraphics[width=0.95 \columnwidth]{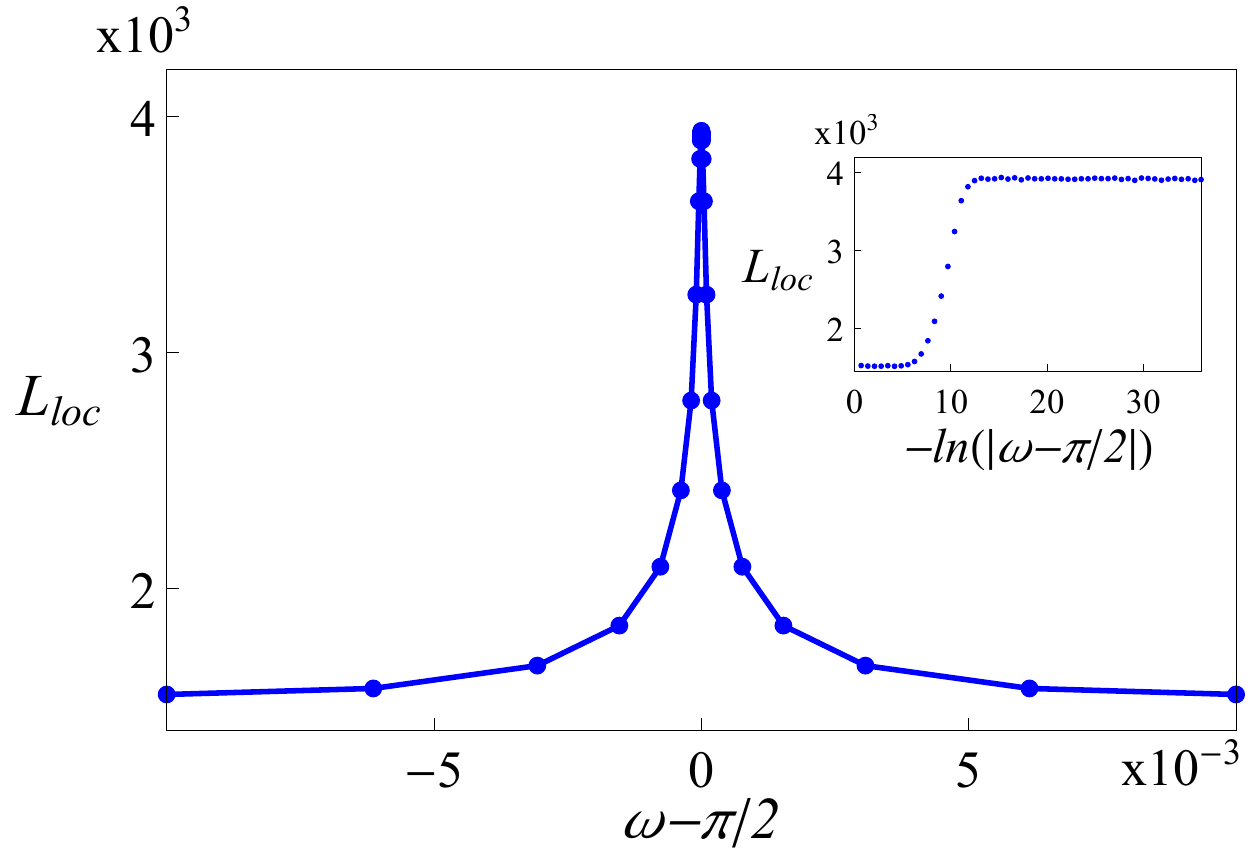}
\caption{The localization length for weak disorder in $\varphi_2$ near the band center $\omega=\pi/2$ ($\theta=\pi/4$).
Here $W=\pi / 50$. Symbols - results of computations, lines are guiding the eye.
Inset: same but resolving the frequency dependence of the localization length on a logarithmic scale close to the band center.}
\label{fig5}
\end{figure}

We remind that the angle $\varphi_2$ can be interpreted as an internal synthetic flux. Without loss of generality we can take $x_0=0$ in (\ref{PDF}).
We observe that the localization length $L_{loc}$ is always finite for any finite but nonzero strength of the disorder $W$, see Fig.\ref{fig4}.
For weak disorder $W \ll \pi$ the localization length
$L_{loc}$ is almost independent of $\omega$ inside the bands of the ordered case, with a small peak in the center of each band ($\omega \simeq \pi/2$).
The localization length inside this peak can double tis value as compared to the plateau values outside the peak, see Fig.\ref{fig5}.
However, according to our computations, the localization length stays finite at the peak center for finite disorder strength (inset Fig.\ref{fig5}).
For strong disorder $W=\pi$ the localization length is frequency independent, see Fig.\ref{fig4}.
Variation of $\theta$ does not qualitatively changes the outcome. However, in the special case $\theta = \pm \pi/2$ the localization length vanishes $L_{loc}=0$.
The eigenstates are then still compactly localized in full accord with (\ref{CLS}), while the eigenfrequencies simply become $\omega_n=\pm \pi/2+\varphi$.

\subsection{Disorder in $\theta$}

\begin{figure}
\includegraphics[width=0.95 \columnwidth]{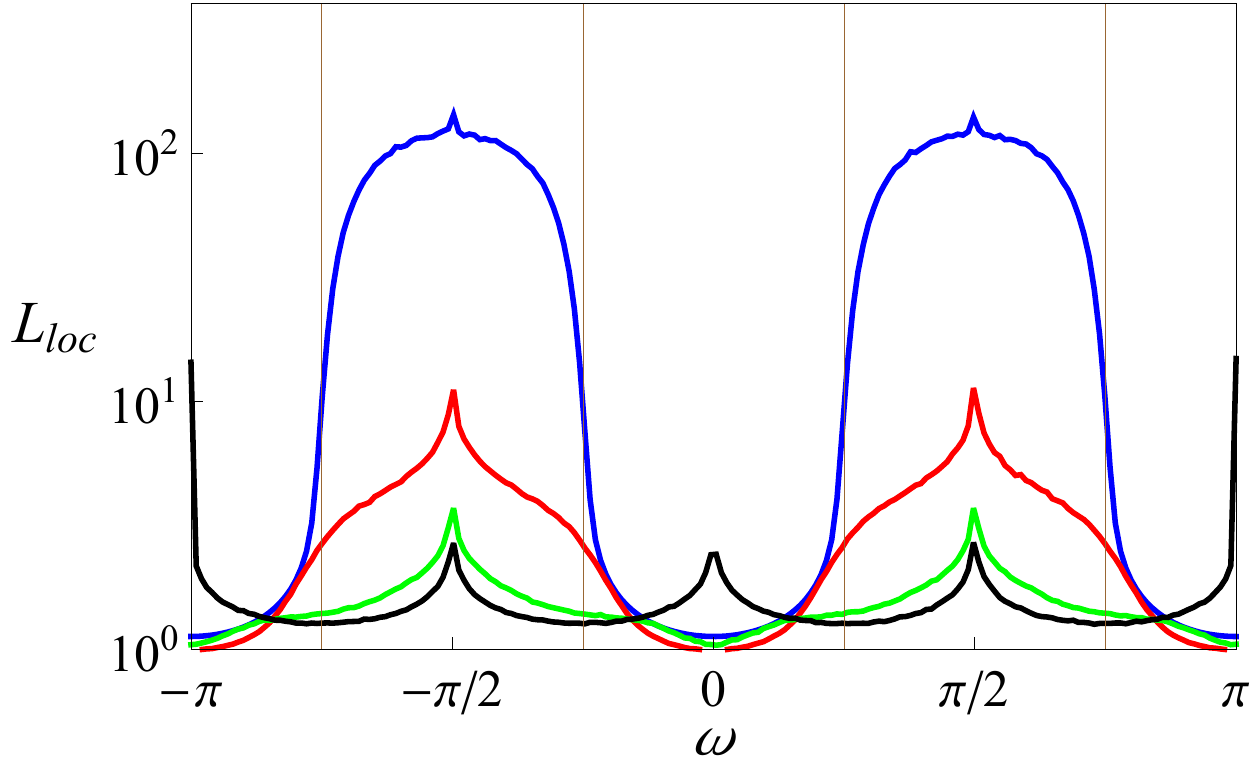}
\caption{
The dependence of the localization length on the frequency $\omega$ for disorder in $\theta$.
From top to bottom at $\omega=\pi/2$: $W=\pi/20,\pi/5,\pi/2,\pi$. Here $\theta=\pi/4$.
The red vertical lines indicate the boundaries of the allowed bands (\ref{Dispersion}).}
\label{fig6}
\end{figure}

\begin{figure}
\includegraphics[width=0.95 \columnwidth]{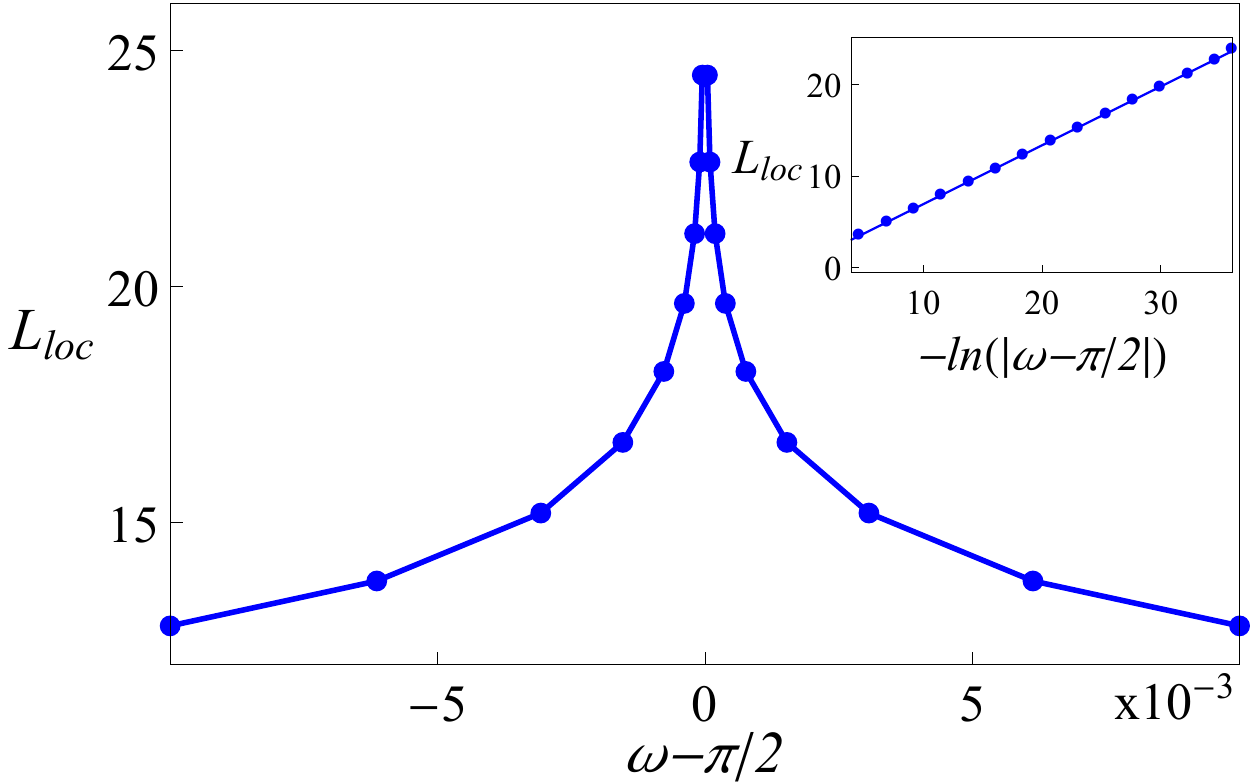}
\caption{
The numerically calculated dependence of the localization length $L_{loc}$ on the characteristic frequency $\omega$ near the band center $\omega=\pi/2$ for disorder in $\theta$.
The average angle $\theta_0=\pi/4$ and the strength of disorder $W=\pi/2$.
Symbols - results of numerical computations, lines guide the eye.
Inset: resolving the frequency dependence of the localization length on a logarithmic scale close to the band center.
Symbols - results of numerical computations. The straight line is a linear fit of the data.
}
\label{fig7}
\end{figure}

\begin{figure}
\includegraphics[width=0.95 \columnwidth]{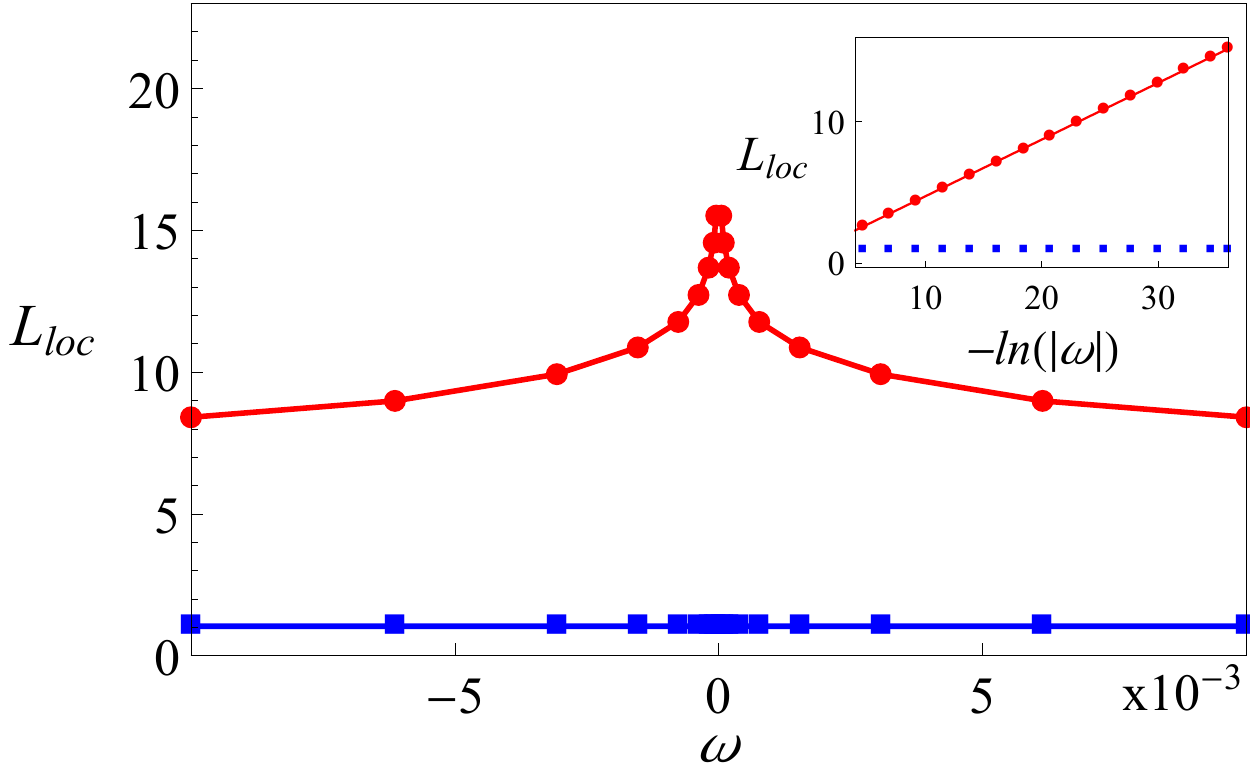}
\caption{
The numerically calculated dependence of the localization length on the frequency close to $\omega=0$ for disorder in $\theta$.
The strength of disorder $W=\pi/2$, and the average angles $\theta_0=0$ (red circles) and $\theta_0=\pi/4$ (blue squares).
Symbols - results of numerical computations, lines guide the eye.
Inset: same but resolving the frequency dependence of the localization length on a logarithmic scale close to $\omega=0$.
The straight line is a linear fit of the data.
}
\label{fig8}
\end{figure}

We remind that the angle $\theta$ can be interpreted as a kinetic energy of a quantum particle, which controls the band width.
At variance to the previous cases, the localization length will diverge logarithmically at the band centers $\omega=\pm \pi/2$ \cite{obuse2011topological}, and results in general depend on the
average $x_0=\theta_0$ in (\ref{PDF}).
While the divergence is barely seen in Fig.\ref{fig6},
it becomes evident in the zoom in Fig.\ref{fig7}(a).
A further logarithmic divergence of the localization length is observed at $\omega=0,\pi$ for the special case of zero average $\theta_0=0$, see Fig.\ref{fig8}.

\section{Analytical results on the localization length}
\label{IV}

In order to provide the analytical results for the localization length $L_{loc}$ for different types of disorder, we use methods of stochastic equations for the phase and amplitude of the
wave function which has been previously used successfully for a one-dimensional tight-binding model with diagonal and off-diagonal disorder  \cite{LGP,Schmidt,Ovchinnikov}. These methods allow us
to systematically study the dependence of the localization length on all relevant parameters.

\subsection{Weak disorder}

For weak disorder $W \ll \pi$
we rewrite the transfer matrix in the following form: $\hat T=\hat T_0+\hat T_{d}$, where the matrix $\hat T_0$ is the transfer matrix of the discrete time quantum walk in the absence of disorder. $\hat T_0$ contains the average angle $\theta_0$, and the average values of the angles $\varphi, \varphi_2$ can be zeroed without loss of generality.
The matrix $\hat T_{d}$ randomly changes from site to site.

As a next step we choose a basis in which $\hat T_0$ is diagonal.  The corresponding unitary matrix
\begin{equation} \label{Rotationmatrix}
\hat S= \left ( \begin{array}{cc}
\tan \theta_0 & i \xi \\
-i\xi & \tan\theta_0\\
\end{array} \right ) ,
\end{equation}
where $\xi=\sin \omega \sec \theta_0-\sin k$, and $k$ is determined by dispersion relationship for a fixed value of $\omega$ in (\ref{Dispersion}). In the new basis the transfer matrix $\hat T_0 $ is written as
 \begin{equation} \label{TransfermatrixHom}
\hat{\tilde T}_0= \hat S^{-1} \hat T_0 \hat S=\left ( \begin{array}{cc}
e^{ik}& 0 \\
0& e^{-ik}\\
\end{array} \right ).
\end{equation}
Similarly the disorder-dependent part of the transfer matrix $T_{d}$ results in $\hat{\tilde T}_{d}= \hat S^{-1} \hat T_{d} \hat S $:
 \begin{equation} \label{Transfermatrix-DisorderGen}
\hat{\tilde T}_{d}= \left ( \begin{array}{cc}
\alpha_n & \beta_n \\
\beta^ \ast_n& \alpha^ \ast_n\\
\end{array} \right ),
\end{equation}
where the parameters $\alpha_n$ and $\beta_n$ are functions of the random quantum coin angles and the average $\theta_0$.

We obtain a stochastic equation for the wave function $\tilde{\Psi}_{+,n}$:
\begin{equation} \label{WavefunctionEq}
\tilde{\Psi}_{+,n}=[e^{ik}+\alpha_n]\tilde{\Psi}_{+,n-1}+\beta_n \tilde{\Psi}^\ast_{+,n-1}\;.
\end{equation}
Introducing the amplitude $r_n$ and phase $\chi_n$ of the wave function $\tilde{\Psi}_{+,n}$ as $\tilde{\Psi}_{+,n}=r_n e^{i\chi_n}$, we arrive at
\begin{equation} \label{WavefunctionEq-2}
\frac{r_n}{r_{n-1}}e^{i[\chi_n-\chi_{n-1}-k]}=1+e^{-ik}\alpha_n+e^{i[-2\chi_{n-1}-k]}\beta_n.
\end{equation}
Thus, if the frequency $\omega$ is located inside of the frequency band gap ( see Fig.\ref{fig2}) the corresponding wave vector $k$ takes an imaginary value, and therefore, one can conclude that  the localization length $L_{loc}$ is bounded from above by $1/|Im (k)|$.

If the frequency $\omega$ is located in the allowed frequency range of the ordered case, the wave vectors $k$ take real values.
For weak disorder $W \ll \pi$ the values of $\alpha_n$ and $\beta_n$ are small and of the order of $W$.
Then it follows that $r_n$ and $(\chi_n-\chi_{n-1})$ vary  weakly from site to site.
Replacing the discrete site variable $n$ by a continuous variable $u$ and replacing differences by differentials , e.g. $r_{n-1} \rightarrow r(u) - dr/du$,
we arrive at the following differential equations:
\begin{widetext}
\begin{equation} \label{DiffEquation}
 \begin{array}{cc}
\frac{d (ln r)}{du} = Re[ \alpha(u)] \cos (k)+Im [ \alpha (u)] \sin(k)
+Re [\beta(u)] \cos(2\chi+k)+Im [\beta(u)] \sin(2\chi+k) \;,\\ \\
\frac{d \chi}{du} = k-Re [\alpha(u)] \sin (k)+Im [\alpha (u)] \cos (k)
+Im [\beta(u)] \cos(2\chi+k)-Re [\beta(u)] \sin(2\chi+k) \; .
\end{array}
\end{equation}
\end{widetext}
For uncorrelated disorder, we solve Eqs. (\ref{DiffEquation})
by using a standard perturbation analysis. In particular, we integrate the second equation in (\ref{DiffEquation}), insert the result into
the first equation, expand up to second order terms in $\alpha(u)$ and $\beta(u)$ and discard fast oscillating terms.
After a final averaging over disorder we obtain an exponential increase of the amplitude of wave function, $<\ln(r)>=u/L_{loc}$ with
the localization length
\begin{equation} \label{Localizationlength}
L_{loc}=\frac{4}{<|\beta(u)|^2>}.
\end{equation}
Here $<|\beta(u)|^2> \equiv N^{-1}\lim_{N \rightarrow \infty} \sum_{n=1}^N |\beta_n|^2$.
Note that the perturbation analysis and Eq.(\ref{Localizationlength}) are not valid if the wave vector $k$ is close to the special points $k=0, \pm \pi/2, \pm \pi$.

\subsubsection{Disorder in $\varphi$}
For disorder in $\varphi$, the random transfer matrix $\hat T_{d}$ takes a diagonal form:
 \begin{equation} \label{TransfermatrixVarphi2}
\hat{T}_{d}= \sec \theta_0  \left ( \begin{array}{cc}
e^{i\omega} [e^{i\varphi_n} -1]& 0\\
0 & e^{-i\omega} [e^{-i\varphi_n} -1]\\
\end{array} \right ).
\end{equation}
Rotating this matrix to the new basis we obtain the parameter $\beta_n$ as
$$
\beta_n=\frac{\tan \theta_0 }{\sin k}[\cos k \sin \varphi_n-\sin \omega \sec \theta_0 (1-\cos \varphi_n)]\;.
$$
With Eq.(\ref{Localizationlength}) this leads to the final result
\begin{equation} \label{Localizationlength-varphi}
L_{loc}=\frac{4 \sin^2 (k) \cot^2(\theta_0)}{ (W^2/3)\cos^2 k +\sin^2 \omega \sec^2 \theta_0 (W^4/20)}.
\end{equation}
We obtain that the localization length $L_{loc} \sim 1/W^2$. However, for $\omega=\pm \pi/2$ this scaling is replaced by
$L_{loc} \sim 1/W^4$, which leads to a strong enhancement of the localization length.
This is the explanation for the observed anomalous enhancement of the localization length in Fig.\ref{fig3}.
In addition, the special gapless case
$\theta_0=0,\pi$ yields complete delocalization $L_{loc} \rightarrow \infty$, as can be also easily observed from the
original equations (\ref{HoppingeqEX},\ref{HoppingeqEX-2}).

\begin{figure}[tbp]
\includegraphics[width=0.95 \columnwidth]{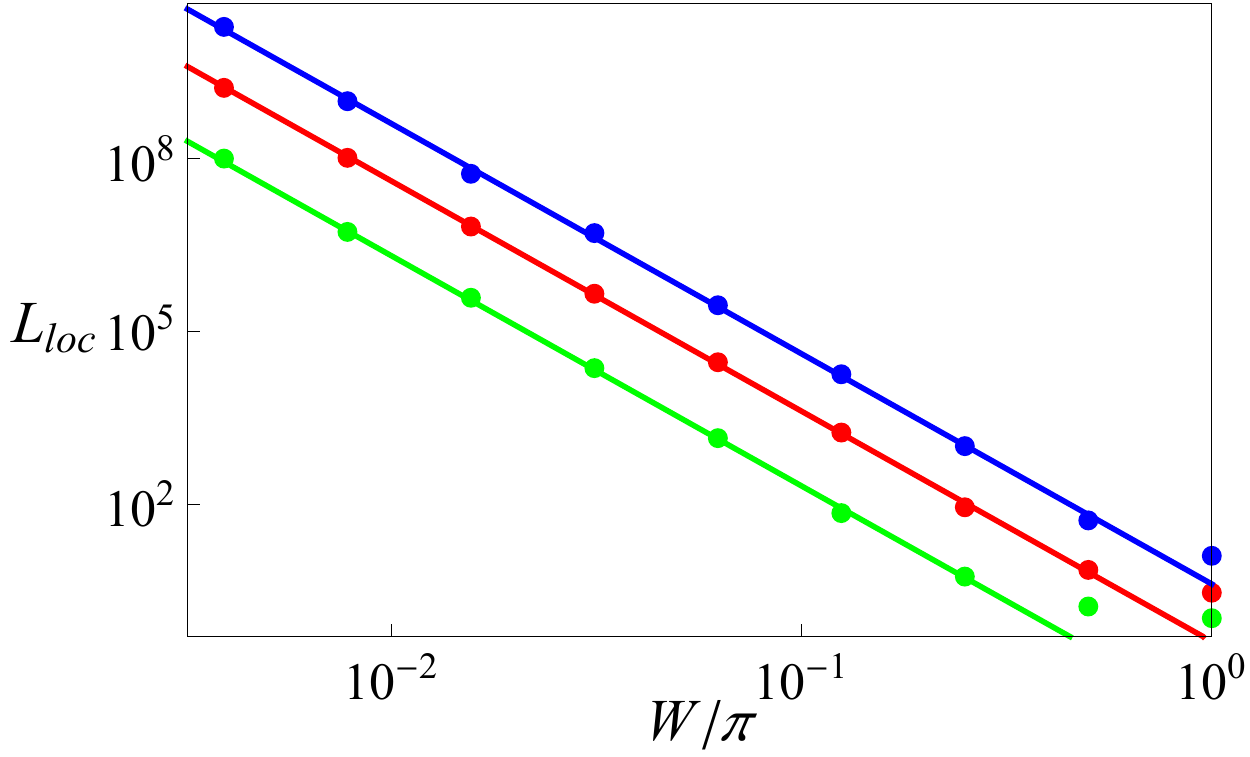}
\caption{The localization length as a function of the disorder strength for disorder in $\varphi$ for $\omega=\pi/2$ (band center). Solid lines - the analytical result (\ref{Localizationlength-varphi}), symbols - numerical computations,
Here $\theta_0 = \pi/8,\pi/4,3\pi/8$ from top to bottom. The predicted scaling $L_{loc} \sim 1/W^4$ is observed.
}
\label{fig9}
\end{figure}

These features are in a good agreement with the numerical computations from the previous section (see Fig. 3).
In particular,
the analytical result (\ref{Localizationlength-varphi}) is in excellent agreement with the computed dependency of $L_{loc}(\omega=\pi/2)$ on the strength of disorder for different values of $\theta$, as shown in
Fig.\ref{fig9}.

\begin{figure}[tbp]
\includegraphics[width=0.95 \columnwidth]{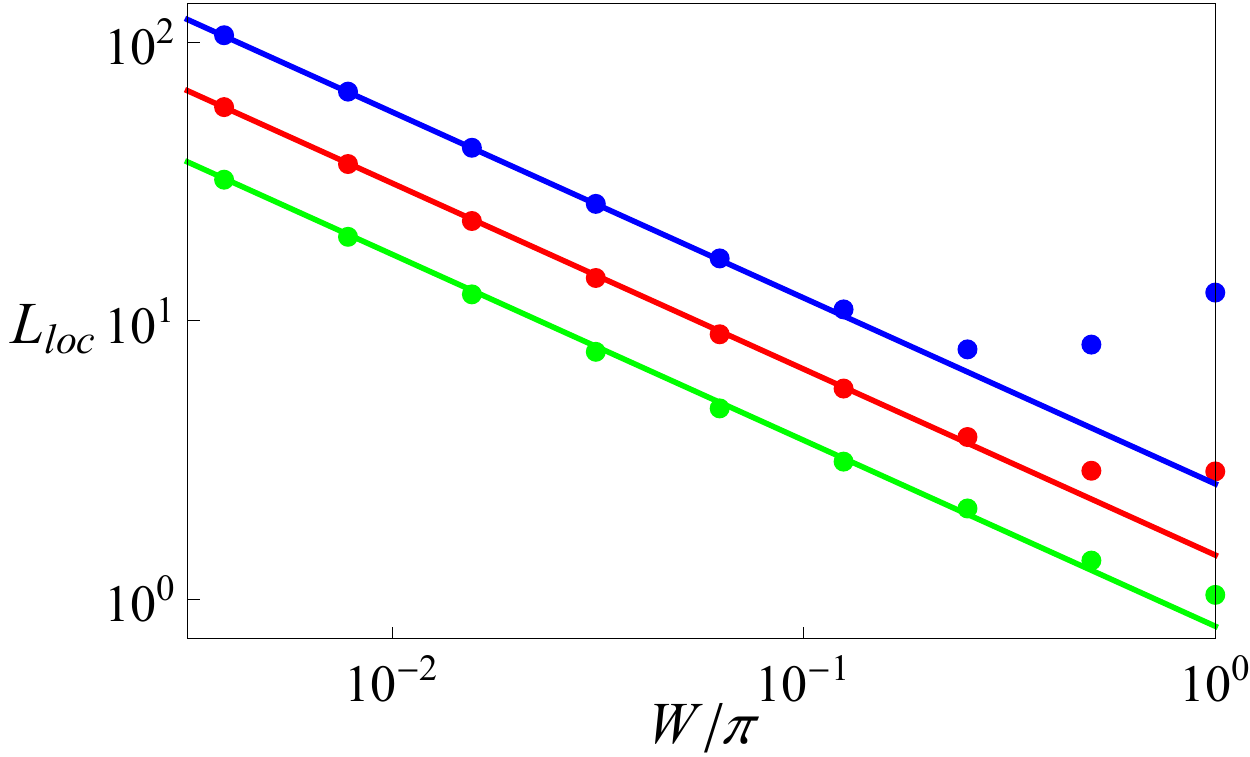}
\caption{The localization length as a function of the disorder strength for disorder in $\varphi$ for $\omega=\theta_0$ (band edge).
Solid lines - the analytical result (\ref{phi-boundary}) with $\eta = 1.36$, symbols - numerical computations.
Here $\theta_0 = \pi/8,\pi/4,3\pi/8$ from top to bottom. The predicted scaling $L_{loc} \sim 1/W^{2/3}$ is observed.
}
\label{fig10}
\end{figure}

We proceed with estimating the localization length on the boundaries of the spectrum $\omega(k)$ of the ordered case, by choosing e.g.
the limit $k \ll 1$ ($\omega \approx \theta_0$).
Using (\ref{Localizationlength-varphi}) we obtain $L_{loc} = 12 k^2/(\tan^2 \theta_0 W^2)$. On the other hand,
for $\omega$ values located inside of gap, $L_{loc} \simeq 1/|k|$.
Both equations can be satisfied by the following scaling of
the localization length on the boundaries of spectrum:
\begin{equation}
L_{loc} = \eta \tan^{-2/3} (\theta_0) W^{-2/3} \;,
\label{phi-boundary}
\end{equation}
where $\eta$ is an unknown prefactor of order one.
In Fig.\ref{fig10} we compare the numerically calculated dependence of $L_{loc}(\omega=\theta)$ on $W$
with the analytical prediction (\ref{phi-boundary}) for various values of $\theta_0$. We find excellent agreement with just one
fitting parameter $\eta = 1.36$ for all cases.

\subsubsection{Disorder in $\varphi_2$}
\begin{figure}[tbp]
\includegraphics[width=0.95 \columnwidth]{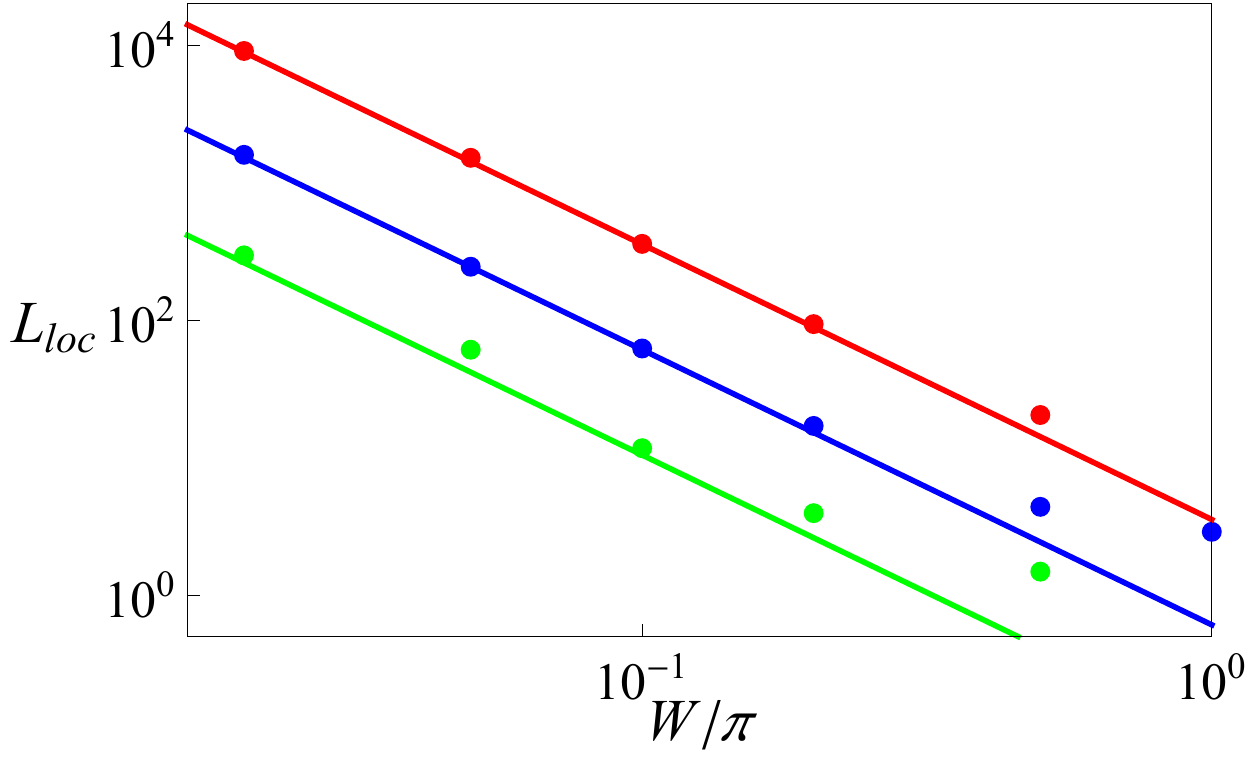}
\caption{
The mean localization length inside the band as a function of the disorder strength for disorder in $\varphi_2$. Solid lines - the analytical result (\ref{Localizationlength-varphi2}) with an additional fitting parameter $\eta=0.5$, symbols - numerical computations.
Here $\theta_0 = \pi/8,\pi/4,3\pi/8$ from top to bottom.
The predicted scaling $L_{loc} \sim 1/W^2$ is nicely observed.
}
\label{fig11}
\end{figure}

For disorder in $\varphi_2$ the random transfer matrix $\hat T_{d}$ has only nonzero off-diagonal terms
 \begin{equation} \label{Transfermatrixvarphi2}
\hat{T}_{d}= \tan \theta_0 \left ( \begin{array}{cc}
0& e^{i\varphi_{2,n}}-1\\
e^{-i\varphi_{2,n}}-1& 0\\
\end{array} \right ).
\end{equation}
Rotating this matrix to the new basis we obtain the parameter $\beta_n$ as
$$
\beta_n=-i \varphi_{2,n}\tan \theta_0 \;.
$$
With Eq.(\ref{Localizationlength})
this leads to the final result
\begin{equation} \label{Localizationlength-varphi2}
L_{loc}=\frac{12}{\tan^2 \theta_0 W^2} \;.
\end{equation}
The localization length is \textit{independent} of  $\omega$ for frequencies $\omega$ inside the bands,
which explains the observed plateaus in Fig.\ref{fig4}.
The predicted scaling with $W$ and $\theta_0$ is in excellent agreement with computational results as shown in Fig.\ref{fig11}, with numerical prefactor being $6$ instead of $12$.

\subsubsection{Disorder in $\theta$}
For disorder in $\theta$ the random transfer matrix $\hat T_{d}$ takes the following form:
 \begin{equation} \label{Transfermatrix-theta}
\hat{T}_{d}= \frac{\theta_n}{\cos^2 \theta_n}\left ( \begin{array}{cc}
e^{i\omega}\sin \theta_0 & 1\\
1& e^{-i\omega}\sin \theta_0\\
\end{array} \right ).
\end{equation}
Rotating this matrix to the new basis we obtain the parameter $\beta_n$  as
$$
\beta_n=\frac{\sin  \omega}{\sin k \cos \theta_0}\theta_n \;.
$$
With Eq.(\ref{Localizationlength})
this leads to the result
\begin{equation} \label{Localizationlength-theta}
L_{loc}=\frac{12\sin^2 k \cos^2 \theta_0}{W^2 \sin^2 \omega}.
\end{equation}
We find that the localization length scales as $L_{loc}\sim 1/W^2$ similar to the previous cases. At the band edge
$k \simeq 0$ ($\omega \simeq \pm \theta$) the localization length scales similar to the case of $\varphi$ disorder as
$L_{loc} \simeq (\tan^2 \theta_0 W^2)^{-1/3}$.

\begin{figure}[tbp]
\includegraphics[width=0.95 \columnwidth]{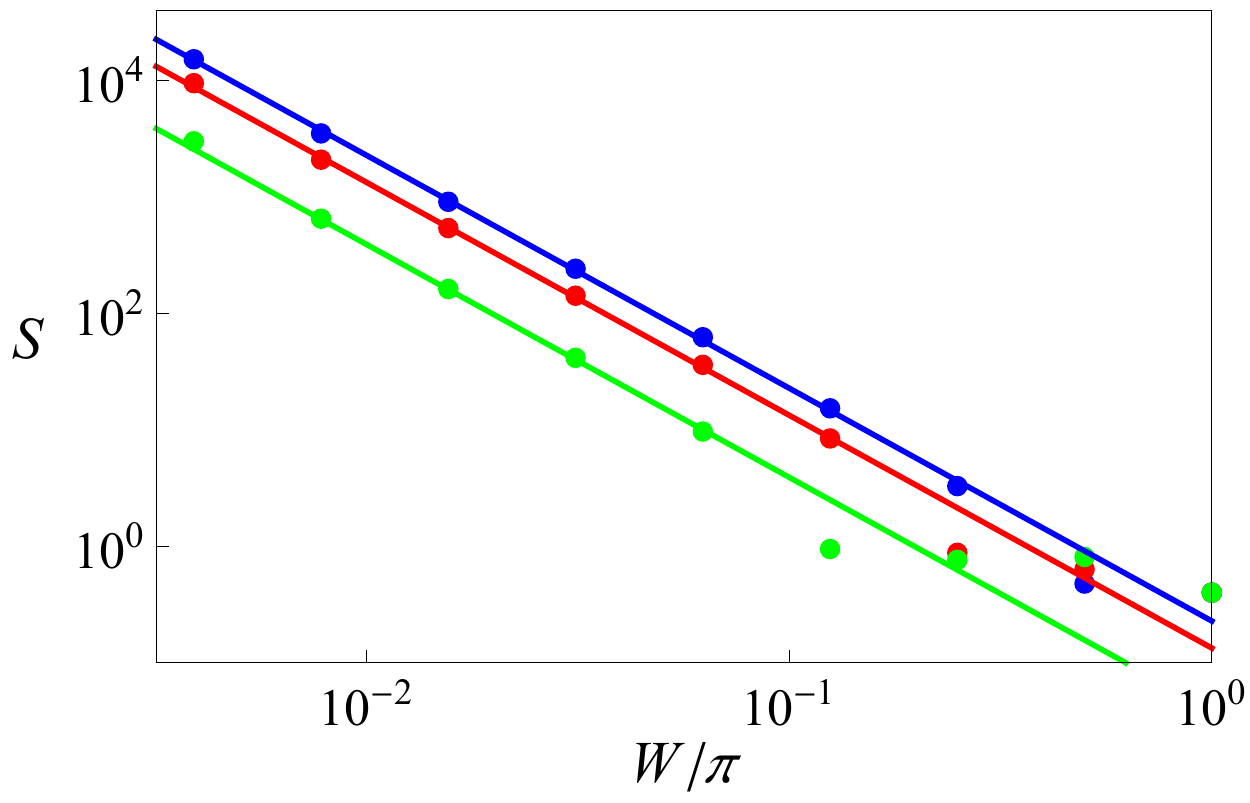}
\caption{
The numerically (symbols) and analytically (solid lines) calculated dependence of the coefficient
$S = \eta \frac{12 \cos^2 \theta_0}{W^2}$  (see Eq. (\ref{Sing-PI/2})) on the strength of disorder $W$, with the numerical fitting parameter $\eta=0.22$.
Here $\theta_0 = \pi/8,\pi/4,3\pi/8$ from top to bottom. The predicted scaling $L_{loc} \sim 1/W^2$ is observed.
}
\label{fig12}
\end{figure}

However, at the band center we observed a logarithmic divergence of the localization length from numerical computations, see Fig.\ref{fig7}.
The divergence of the localization length at the precise band center was derived in Ref.\onlinecite{Singul}.
This follows from the fact, that the parameter $\alpha_n$ in the Eq. (\ref{Transfermatrix-DisorderGen}) is  strictly vanishing
at the band center. Therefore $\alpha_n$ is a higher order perturbation term and can be neglected close to the band center as well.
Eq. (\ref{WavefunctionEq}) is then reduced to
\begin{equation} \label{WaveFunctTransfer-PI/2}
\tilde{\Psi}_{+,n}=e^{ik}\tilde{\Psi}_{+,n-1}+\beta_n \tilde{\Psi}^\ast_{+,n-1}.
\end{equation}
The corresponding differential equations (\ref{DiffEquation})
modify into
\begin{equation} \label{DiffEquation-PI/2}
 \begin{array}{cc}
\frac{d (ln r)}{du} =& \beta(u) \sin[2\chi] \;, \\
&\\
\frac{d \chi}{du} =& \pi/2+\delta \omega-\beta(u) \cos[2\chi] \;.
\end{array}
\end{equation}
Excluding $\beta(u)$ we find $<\ln r>=<\delta \omega \tan(2\chi)>u$, and therefore, the corresponding localization length is $L_{loc} =[<\delta \omega \tan (2\chi)>]^{-1}$. In order to compute the average, we  introduce a new variable $z=2\ln [\tan(\chi-\pi/4)]$
and rewrite the second equation in (\ref{DiffEquation-PI/2}) as
\begin{equation} \label{WaveFunctTransfer-PI/2-new}
\frac{dz}{du}=4(\delta \omega) \cosh z+4\beta(u),
\end{equation}
with $L_{loc}=[<(\delta \omega) \sinh (z/2)>]^{-1}$. In order to find the average value of $z$ we transfer from the stochastic Eq. (\ref{WaveFunctTransfer-PI/2-new}) to the corresponding Fokker-Planck equation for the probability $P(z)$, which satisfies
\begin{equation} \label{Fokker-Planck}
\frac{16 W^2}{\cos^2 \theta_0}\frac{d^2 P(z)}{dz^2}-4(\delta \omega) \frac{d}{dz}[\cosh (z/2) P(z)]=0
\end{equation}
with the normalization condition $\int_0^{2\pi} d\chi P(z)=1$. It follows that  ${L_{loc}}^{-1}= \delta \omega \int dz \sinh (z/2) P(z)$. In the limit $\delta \omega \ll W^2$ we obtain a logarithmic enhancement of the localization length as
\begin{equation} \label{Sing-PI/2}
L_{loc}=S \ln |\frac{\delta \omega}{W^2}|\;,\; S = \frac{12 \cos^2 \theta_0}{W^2} \;.
\end{equation}
This dependence on $\theta_0$ and $W$ agrees excellently with numerical data in Fig.\ref{fig12}, with a numerical prefactor being different. Notice here that this logarithmical divergence  resembles a well-known Dyson-Wigner singularity obtained previously in the electronic transport of one-dimensional disordered tight-binding chain in the presence off-diagonal disorder \cite{LGP,Schmidt,Ovchinnikov,DW}.

\subsection{Strong disorder}
We start with noting that the linear transfer matrix equations (\ref{TransfermatrixEq}) which define a linear two-dimensional map,
can be equivalently rewritten as a one-dimensional map, which is however nonlinear.
We introduce the variable $y_n$
\begin{equation}
    y_n = \frac{\hat{\Psi}_{+,n}}{\hat{\Psi}_{-,_n}},
\end{equation}
in order to rewrite the two-component wave function $\hat{\Psi}_n $ as
\begin{equation}
\hat{\Psi}_n = A_n \begin{pmatrix}
y_n \\ 1
\end{pmatrix} \;.
\end{equation}
This definition allows one to reduce the two-dimensional map defined by the transfer matrix (\ref{Trmatrix})
to a one-dimensional map:
\begin{equation}\label{initial_m_map}
\begin{split}
y_{n+1} &= M(y_n) = e^{2i \varphi_{2,n}} \frac{c_n e^{i \lambda_n} y_n + 1}{y_n + c_n e^{-i\lambda_n}}, \\ c &= 1 / \sin(\theta_n), \; \lambda_n = \omega + \varphi_n - \varphi_{2,n}.
\end{split}
\end{equation}

The complex variable $y_n$ takes random values, and is characterized by a stationary probability distribution $P(y)$\cite{derrida1983singular, crisanti2012products}. Taking into account that the absolute value of the two-component wave function shows an exponential increase as $n$ goes to infinity, we obtain the  localization length $L_{loc}$  as
\begin{equation}
1/L_{loc} = \lim_{N \rightarrow \infty}
\frac{1}{N}\sum_{n=1}^{N} \ln \left(\left| \hat{\Psi}_{n+1} \right|/\left|\hat{\Psi}_n\right|\right).
\end{equation}
By making use of (\ref{TransfermatrixEq}) we obtain
\begin{widetext}
\begin{equation}\label{general_localization_length}
    1/L_{loc} = \left< \ln \left( \left| \hat{T} \hat{\Psi} \right| / \left|  \hat{\Psi} \right| \right) \right>
   = \int \diff \mu(\zeta) \int_{0}^{\infty}\diff \rho \int_{-\pi}^{\pi} \diff \phi \; P(\rho e^{i \phi}) \cdot \ln \left(
    \left|
    \hat{T}(\zeta) \left(
\begin{array}{c}
    \rho e^{i \phi} \\
    1
\end{array}
\right)
    \right| / \sqrt{1+\rho^2}
    \right),
\end{equation}
\end{widetext}
where we define $y=\rho e^{i \phi}$ and $\zeta$ is a random angle ($\varphi$, $\varphi_2$, or $\theta$), and $\mu(\zeta)$ is its measure.

The map (\ref{initial_m_map}) reduces the absolute value $\rho$ if $\rho>1$ and increases it if $\rho<1$.
Thus $\rho=1$ for the stationary distribution of $P(y)$. Then $P(y)$ has the following form:
\begin{equation}\label{delta_rho_1_distribution}
P(y) = p(\phi) \delta(\rho - 1).
\end{equation}
The dynamics of the phase $\phi_n$ (which is defined  $\mathrm{mod} \; 2 \pi$) is determined by the following stochastic equation:
\begin{equation}\label{phi_map}
\begin{split}
\phi_{n+1} = m(\phi_n)=2 \arg(\kappa(\lambda_n, \phi_n)) + 2 \varphi_{2,n}- \phi_n  &\;, \\
\kappa(\lambda_n, \phi_n) = \sin(\theta_n)+e^{i(\lambda_n+\phi_n)} &\;.
\end{split}
\end{equation}
This dynamic equation is reduced to an integral equation for the distribution $p(\phi)$:
\begin{equation}\label{stat_distr_equation_lambda}
p(\phi^\prime) = \int_{<\zeta> - W}^{<\zeta> + W} \frac{\diff \zeta}{2W} \int_{-\pi}^{\pi} \diff \phi p(\phi) \delta(\phi^\prime - m(\phi)) \;.
\end{equation}

\subsubsection{Disorder in $\varphi$ and $\varphi_2$}

\begin{figure}
\includegraphics[width=0.95 \columnwidth]{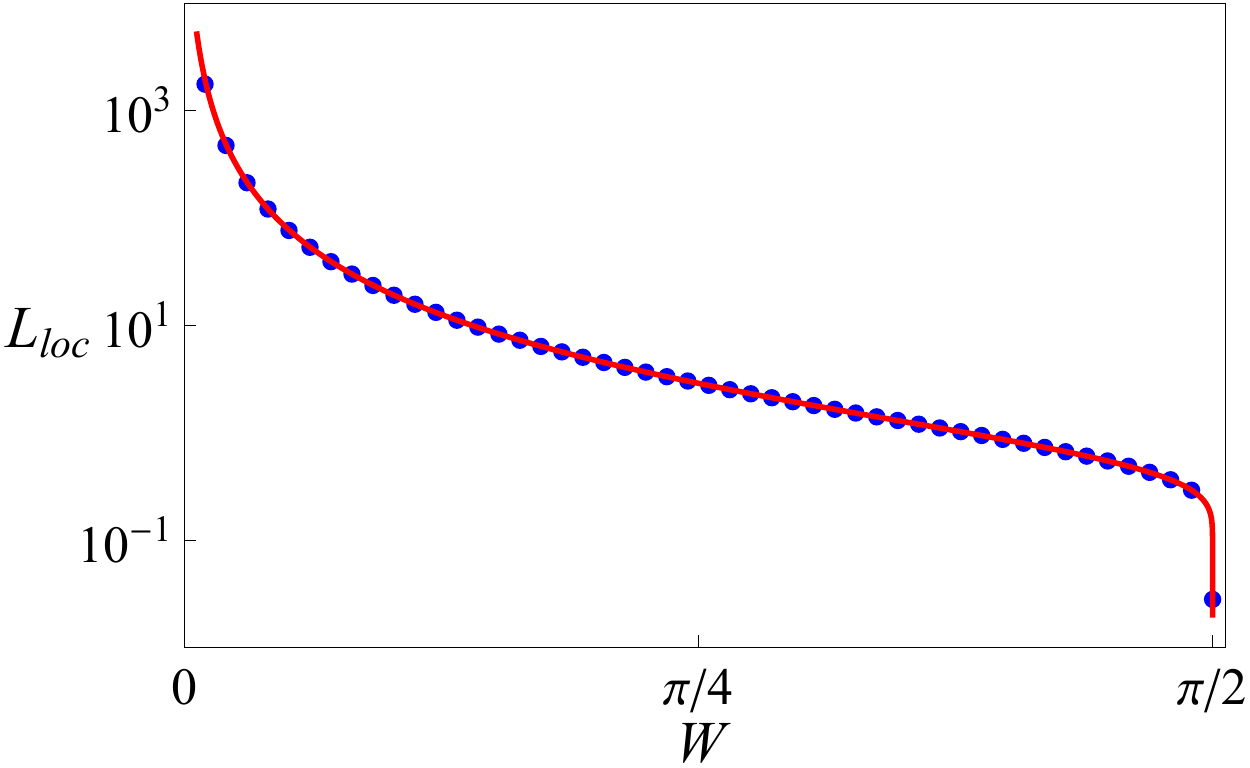}
\caption{Localization length under maximal disorder in $\phi$. Red solid lines is the analytical result (\ref{LocLength-MD}), blue dots are numerical.
}
\label{fig13}
\end{figure}

We first consider disorder in $\varphi$.
We present results for the case of strongest disorder $W=\pi$.
As $\omega$ only appears in combination $\omega + \varphi$, and integration in (\ref{stat_distr_equation_lambda}) is over the whole period in this case, one may disregard $\omega$ by shifting variables. $\phi_2$ is a fixed constant, which allows to eliminate it in a similar way. This yields
\begin{equation}
p(\phi^\prime) = \dfrac{1}{2 \pi}\int^\pi_{-\pi} \diff \lambda \; p \left(2 \arg(\kappa(\lambda, 0)) - \phi^\prime \;\right).
\end{equation}
This equation is satisfied by the uniform solution
\begin{equation}\label{uniform_phi_distrib_solution}
p(\phi)=\frac{1}{2\pi}.
\end{equation}
Substituting (\ref{delta_rho_1_distribution}) and (\ref{uniform_phi_distrib_solution}) with the transfer matrix (\ref{Trmatrix}) simplifies (\ref{general_localization_length}),
\begin{equation}
\begin{split}
\frac{1}{L_{loc}} =  \iint_{-\pi}^{\pi} \frac{\diff \varphi \diff \phi}{8\pi^2} \ln \left( \frac{1 + \sin^2\theta +2 \sin\theta\cos(\varphi+\phi)}{ \cos^2\theta}\right).
\end{split}
\end{equation}
Integrating separately over the logarithm of the enumerator (which strictly vanishes) and the denominator we finally arrive at
\begin{equation} \label{LocLength-MD}
L_{loc}=-\frac{1}{\ln(|\cos(\theta)|)}.
\end{equation}
Thus, the localization length is independent of $\omega$ and is determined only by the value of $\theta_0$.
Exactly the same results will hold for strongest disorder in $\varphi_2$ as well, and equation (\ref{LocLength-MD}) again applies.
In Fig.\ref{fig13} we plot the analytical result (\ref{LocLength-MD}) and compare to numerical computations using
the transfer matrix approach, with excellent agreement.

\subsubsection{Disorder in $\theta$}
\begin{figure}
\includegraphics[width=0.95 \columnwidth]{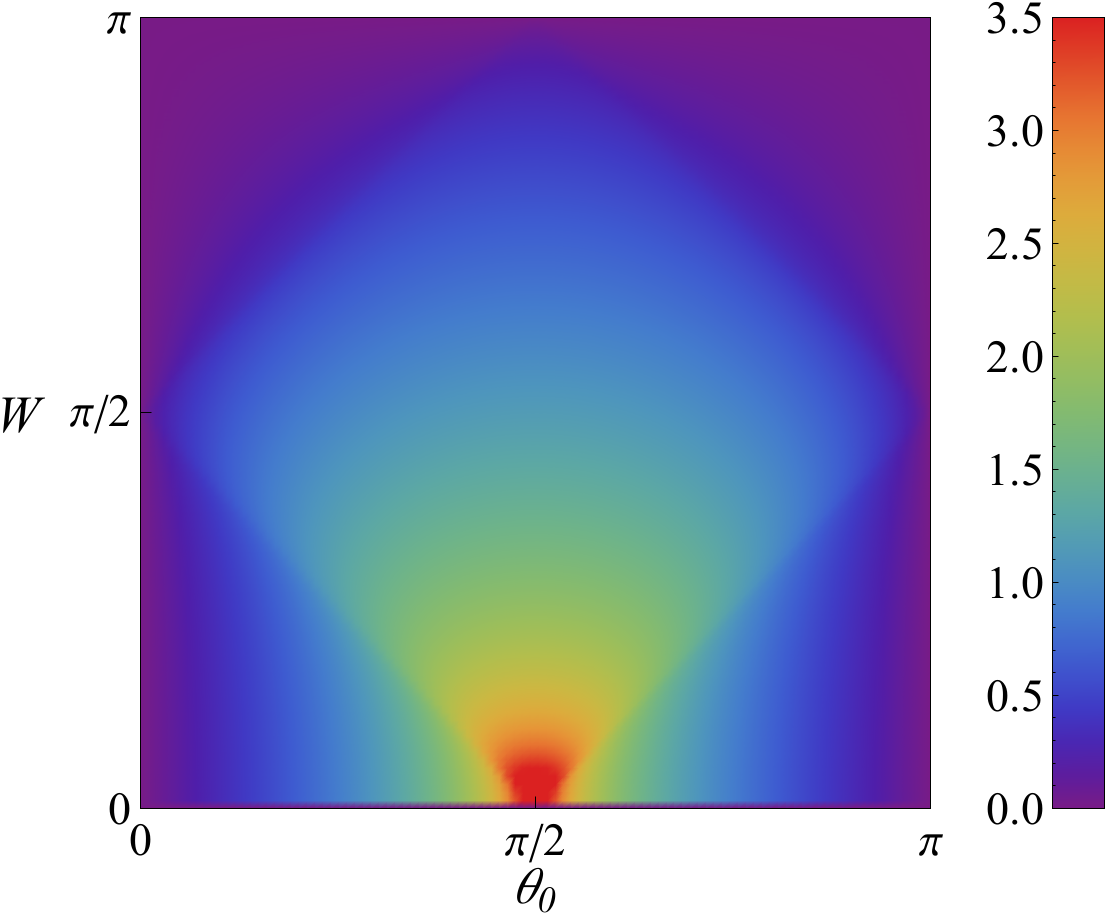}
\caption{The inverse localization length $1/L_{loc}(\omega=0)$ as a function of  $\theta_0$ and $W$ according to
Eq. (\ref{omega0_local_length}). The inverse localization length strictly vanishes on the lines $\theta=0,\pi$ and $W=\pi$.}
\label{fig14}
\end{figure}

In this subsection we analyze the singular behavior of $L_{loc}(\omega=0)$ and $L_{loc}(\omega=\pm \pi/2)$ for disorder in $\theta$.
We start with $\omega=0$ (without loss of generality we choose $\varphi=\varphi_2=0$).
Then, (\ref{phi_map}) reduces to
\begin{equation}
\kappa=\sin(\theta_n) + e^{i \phi} \;.
\end{equation}
It follows that $\phi=0$ is a fixed point of Eq. (\ref{phi_map}), and therefore, $p(\phi) = \delta(\phi)$ solves Eq.(\ref{stat_distr_equation_lambda}). Substituting into (\ref{general_localization_length}) we obtain
\begin{equation}
1/L_{loc} = \frac{1}{2 W} \int_{\theta_0-W}^{\theta_0+W} \diff \theta \ln |cot(\pi/4-\theta/2)|
\nonumber
\end{equation}
to arrive at
\begin{widetext}
\begin{equation}\label{omega0_local_length}
1/L_{loc} =\frac{1}{2 W} \Big| Cl_2\left(\pi/2 + \theta_0 -  W\right) + Cl_2\left(\pi/2 - \theta_0 +  W\right) - Cl_2\left(\pi/2 +  \theta_0 +  W\right) - Cl_2\left(\pi/2 - \theta_0 - W\right) \Big| \;,
\end{equation}
\end{widetext}
where $Cl_2(x)$ is the Clausen function of 2nd order (see Ref.\onlinecite{AbrStegan}). Thus, we obtain \textit{delocalized states} in two particular cases:  either for $ \theta_0 = 0$ with arbitrary disorder strength, or for the case of strongest disorder $W=\pi$ and any value of $\theta_0$,
as shown in Fig. \ref{fig14}.

Next we consider the case $\omega=\pi/2$. Eq. (\ref{phi_map}) is reduced to
\begin{equation}
\kappa=\sin(\theta_n) +e^{i(\pi/2+ \phi)}  \;.
\end{equation}
We find $m(\pi/2)=3\pi/2$, $m(3\pi/2)=\pi/2$ as a period-two limit cycle solution of the  map (\ref{phi_map}).
Thus
\begin{equation}
p(\phi) = \frac{1}{2} \delta(\phi-\pi/2) + \frac{1}{2} \delta(\phi-3\pi/2) \;.
\end{equation}
Substitution into (\ref{general_localization_length}) yields $1/L_{loc} = 0$ for any set of parameters.
We arrive at the result that the localization length strictly diverges at $\omega=\pm \pi/2$, in agreement with our results
for weak disorder Eq. (\ref{Sing-PI/2}).

\section{Conclusion}
\label{V}

In conclusion we have theoretically (numerically and analytically) analyzed the discrete time quantum walk in the presence of spatial disorder. The dynamics of such a quantum walk is determined by four angles of
a quantum coin operator, i.e. $\varphi, \varphi_1, \varphi_2, \theta$ (\ref{coin_operator}). In the absence of spatial disorder the dynamics of the discrete time quantum walk is characterized by the dispersion relation, i.e. the dependence of the characteristic frequency $\omega$ on the wave vector $k$ (\ref{Dispersion}).  The spectrum $\omega(k)$ contains two bands, and is tuned by varying the angle $\theta$. For $\theta=0$ a gapless spectrum occurs, while for $\theta=\pm \pi/2$ the spectrum consists of two gapped flat bands. The equations, the spectrum and the eigenvectors are invariant under two symmetry operations:
bipartite and particle-hole symmetries.

Disorder in the external synthetic gauge field $\varphi_1$ does not impact the extended nature of the eigenstates, and does not destroy the above two symmetries. However, disorder in any of the
remaining three angles $\theta,\varphi,\varphi_2$ enforces Anderson localization of the eigenstates. In particular, disorder in the kinetic energy angle $\theta$ leads to a logarithmic divergence
of the localization length for particular values of the eigenfrequency $\omega$, while again keeping the bipartite and particle-hole symmetries untouched.
Disorder in the onsite energy angle $\varphi$ and the internal synthetic flux angle $\varphi_2$ is destroying the particle-hole symmetry, and yields finite localization length for all allowed eigenfrequencies $\omega$.
Remarkably we obtain that strongest disorder $W=\pi$ in $\varphi$ and $\varphi_2$ yields Anderson localized random eigenstates with a unique localization length, which depends only on $\theta$, but does not
change for different eigenfrequencies $\omega$. This is possible because the space of eigenfequencies is compact and confined to the spectrum of a phase of a complex number residing on the unit circle.

We derive various scaling laws in the limit of weak and strong disorder, and obtain excellent agreement with numerical results using a transfer matrix approach. These results underline the
richness of the considered system, which makes it not only attractive for application reasons, but also an ideal playground for various extensions including the impact of
many body interactions, mean field nonlinearities, and flat band physics, to name a few.

\begin{acknowledgments}
This work was supported by the Institute for Basic Science, Project Code (IBS-R024-D1).
M.V. F. acknowledges the financial support of the Ministry of Education and Science of the Russian Federation in the framework of Increase Competitiveness Program of NUST "MISiS" $K2-2016-067$, and the Russian Science Foundation (grant No. 16-12-00095).
\end{acknowledgments}

\bibliography{QW}

\end{document}